\documentclass[aps,prep,preprintnumbers,amsmath,amssymb,nofootinbib]{revtex4}
\draft
\newcommand {\eps}{\epsilon}
\newcommand {\ga}{\gamma}
\newcommand {\de}{\delta}
\newcommand {\Ga}{\Gamma}

\newcommand {\la}{\lambda}
\newcommand {\al}{\alpha}
\newcommand {\be}{\beta}
\newcommand {\pa}{\partial}
\newcommand {\na}{\nabla}
\newcommand {\fr}{\frac}

\newcommand {\cB}{{\cal B}}

\newcommand {\ch}{{\cal H}}
\newcommand {\cu}{{\cal U}}
\newcommand {\cv}{{\cal V}}

\newcommand {\ct}{{\cal T}}
\newcommand {\beg}{\begin{equation}}
\newcommand {\en}{\end{equation}}
\newcommand {\bega}{\begin{eqnarray}}
\newcommand {\ena}{\end{eqnarray}}
\begin{document}
\title{Semiclassical Gravitoelectromagnetic inflation in a Lorentz gauge: seminal inflaton fluctuations and
electromagnetic fields from a 5D vacuum state}
\author{ $^{1,2}$ Federico Agust\'{\i}n Membiela \footnote{
E-mail address:membiela@mdp.edu.ar}, $^{1,2}$ Mauricio Bellini
\footnote{E-mail address: mbellini@mdp.edu.ar}}
\address{$^{1}$ Departamento de F\'{\i}sica, Facultad de Ciencias Exactas y
Naturales, Universidad Nacional de Mar del Plata, Funes 3350, (7600)
Mar del Plata,
Argentina.\\
$^{2}$ Instituto de F\'{\i}sica de Mar del Plata (IFIMAR), Consejo
Nacional de Investigaciones Cient\'{\i}ficas y T\'ecnicas
(CONICET). }

\begin{abstract}
Using a semiclassical approach to Gravitoelectromagnetic Inflation
(GEMI), we study the origin and evolution of seminal inflaton and
electromagnetic fields in the early inflationary universe from a
5D vacuum state. The difference with other previous works is that
in this one we use a Lorentz gauge. Our formalism is naturally not
conformal invariant on the effective 4D de Sitter metric, which
make possible the super adiabatic amplification of magnetic field
modes during the early inflationary epoch of the universe on
cosmological scales.
\end{abstract}

\maketitle

\section{Introduction}

The origin of cosmological scales magnetic fields is one of the
most important, fascinating and challenging problems in modern
cosmology. Many scenarios have been proposed to explain them.
Magnetic fields are known to be present on various scales of the
universe\cite{3}. Primordial large-scale magnetic fields may be
present and serve as seeds for the magnetic fields in galaxies and
clusters.

Until recently the most accepted idea for the formation of
large-scale magnetic fields was the exponentiation of a seed field
as suggested by Zeldovich and collaborators long time ago. This
seed mechanism is known as galactic dynamo. However, recent
observations have cast serious doubts on this possibility. There
are many reasons to believe that this mechanism cannot be
universal. This is why the mechanism responsible for the origin of
large-scale magnetic fields is looked in the early universe, more
precisely during inflation\cite{turner}, which should be amplified
through the dynamo mechanism after galaxy formation. In principle,
one should be able to follow the evolution of magnetic fields from
their creation as seed fields through to dynamo phase
characteristic of galaxies. It is believed that magnetic fields
can play an important role in the formation and evolution of
galaxies and their clusters, but are probably not essential to our
understanding of large-scale structure in the universe. However,
an understanding of structure formation is paramount to the
problem of galactic and extragalactic magnetic fields\cite{1,2}.

It is natural to look for the possibility of generating
large-scales magnetic fields during inflation with strength
according with observational data on cosmological scales: $<
10^{-9}$ Gauss. However, the FRW universe is conformal flat and
the Maxwell theory is conformal invariant, so that magnetic fields
generated at inflation would come vanishingly small at the end of
the inflationary epoch. The possibility to solve this problem
relies in produce non-trivial magnetic fields in which conformal
invariance to be broken.

On the other hand, the five dimensional model is the simplest
extension of General Relativity (GR), and is widely regarded as
the low-energy limit of models with higher dimensions (such as 10D
supersymmetry and 11D supergravity). Modern versions of 5D GR
abandon the cylinder and compactification conditions used in
original Kaluza-Klein (KK) theories, which caused problems with
the cosmological constant and the masses of particles, and
consider a large extra dimension. In particular, the Induced
Matter Theory (IMT) is based on the assumption that ordinary
matter and physical fields that we can observe in our 4D universe
can be geometrically induced from a 5D Ricci-flat metric with a
space-like noncompact extra dimension on which we define a
physical vacuum\cite{IMT}.

Gravitoelectromagnetic Inflation (GEMI)\cite{gemi} was proposed
recently with the aim to describe, in an unified manner,
electromagnetic, gravitational and the inflaton fields in the
early inflationary universe, from a 5D vacuum. It is known that
conformal invariance must be broken to generate non-trivial
magnetic fields. A very important fact is that in this formalism
conformal invariance is naturally broken. Other conformal symmetry
breaking mechanisms have been proposed so far\cite{tur}. In the
framework of the IMT, electromagnetic effects were studied at a
classical level in\cite{liko}. However, most of these are
developed in the Coulomb gauge. In this paper we study a
semiclassical approach of this formalism in a Lorentz gauge.
Furthermore, for simplicity we shall neglect back reaction effects
on the semiclassical Einstein equations.

\section{Vector fields in 5D vacuum}

We begin considering a 5D manifold $\cal{M}$ described by a
symmetric metric $g_{ab}=g_{ba}$\footnote{In our conventions latin
indices "a,b,c,..,h" run from $0$ to $4$, greek indices run from
$0$ to $3$ and latin indices "i,j,k,..." run from $1$ to $3$.}.
This manifold $\cal M$ is mapped by coordinates $\{x^a\}$
 \beg
    dS^2=g_{ab}dx^a dx^b,\label{meta}
 \en
 where $g_{ab}$ is the 5D tensor metric, such that $g_{ab}=
 g_{ba}$. From the geometrical point of view, to describe a
 relativistic 5D vacuum, we shall consider that $g_{ab}$ is such
 that the Ricci tensor $R_{ab}=0$, and hence: $G_{ab}=0$.
To describe the system we introduce the action on the manifold
$\cal M$
 \beg
    \mathcal{S}=\int d^5x\sqrt{-g}\left[\fr{^{(5)}\,R}{16\pi
    G}-\fr{1}{4}Q_{bc}Q^{bc}\right],
 \en
where $^{(5)}\,R$ is the 5D scalar curvature on the
five-dimensional metric (\ref{meta}) and $Q^{ab}=F^{ab}-\ga
g^{ab}\na_f A^f$, where the 5D Faraday tensor is $F^{bc}=\na^b
A^c-\na^c A^b=\pa^bA^c-\pa^cA^b$. We shall consider that the
fields $A^b$ are minimally coupled to gravity and free of
interactions, so that the second term in the action is purely
kinetic.

\subsection{Einstein Equations in 5D}

If we minimize the action respect to the metric we will obtain
Einstein Equations in 5D. In this paper we shall
 use a semiclassical approach where the Einstein equations are
 expressed by the homogeneous component of the fields. This slightly
 differs from the one used by \cite{Birrel} in the fact that we
 don't need to renormalize the stress tensor, but at the cost of
 assuming a semiclassical behavior of the fields that rules out the
 dependence with the wavenumber in the calculation of the semiclassical
 Einstein equations
  \beg
    G_{ab}=-8\pi G \,T^{(0)}_{ab},
  \en
where $T^{(0)}_{ab}\equiv \langle T_{ab}(\bar{A}^c)\rangle$.
Notice that we use a semiclassical expansion of the vector fields
\begin{equation}
A^c=\bar{A}^c+\de A^c,
\end{equation}
where the overbar symbolizes the 3D spatially homogeneous
background field consistent with the fixed homogeneous metric and
$\de A^c$ describes the fluctuations with respect to $\bar{A}^c$.
In this sense when we perform the expectation value of the stress
tensor, adopting the ansatz $\langle\de A_c\rangle=0$, only will
appear zero order $T^{(0)}_{ab}$ and the second order
$T^{(2)}_{ab}$ in perturbations terms. The last corresponds to a
feedback term and is related to back-reaction effects, which do
not will be consider in this paper.
 The stress tensor is defined by the fields lagrangian being
 symmetric by definition
 \beg
    T_{bc}=\fr{2}{\sqrt{g}}\left\{\fr{\pa}{\pa g^{bc}}\left(\sqrt{g}{\cal L}_f\right)
                                   -\fr{\pa}{\pa x^e}\left[\fr{\pa}{\pa {g^{bc}}_{,\,e}}
                                   \left(\sqrt{g}{\cal
                                   L}_f\right)\right]\right\}.
 \en
The appearance of variations respect to derivatives of the metric
is because we are dealing with vector fields whose covariant
derivative operators involve Christoffel symbols (i.e. ordinary
derivatives of the metric).

 In our case the stress tensor reduces to
 \bega
    T_{bc}&=&{F^e}_b F_{ce}+\fr{1}{4}g_{bc}F_{de}F^{de}
    -\la\left\{2{A^e}_{;\,e}\left[A_{(b;\,c)}-\left(2A^{}_{(b}g^{}_{c)h,\,f}+
    g^{}_{hf,\,(b}A^{}_{a)}\right)g^{hf}\right]+\right.\\
    &+& \left. g_{bc}\left[\left(A^e_{,\,ef}+\Ga^e_{de,\,f}A^d+
    \Ga^e_{de}A^d_{,\,f}+2\Ga^e_{ef}A^d_{,\,d}+\fr{3}{2}\Ga^e_{ed}\Ga^a_{af}A^d
    \right)A^f+\fr{1}{2}\left(A^e_{,\,e}\right)^2\right]\right. \nonumber \\
    &+&  \left. g_{bc,\,f}A^f  A^e_{;\,c}\right\},
 \ena
where $\ga^2=\fr{2\la}{5}$.

\subsection{5D dynamics of the fields}

The Euler-Lagrange equations give us the dynamics for $A_b$

 \beg
    \na_f\na^f A^b-R_f^b A^f-(1-\la)\na^b\na_fA^f=0.
 \en
In particular, the choice $\la=1$ is known as Feynman gauge,
somehow equivalent to a Lorentz gauge $\na_fA^f=0$. In this paper
we shall choose simultaneously both conditions. It is easy to show
that the 5-divergence of the field equation of motions satisfy the
same equation as in a Minkowski space, but changing ordinary
partial derivatives by the covariant derivative
 \beg
    \na^a\na_a\left(\na_f A^f\right)=0.
 \en
Hence, the Lorentz gauge is satisfied for appropriate initial
conditions of $\na_a A^a=0$. With such a choice the field
lagrangian density ${\cal L}_f=-\fr{1}{4}Q^2$ is
 \beg
    {\cal L}'_f=-\fr{1}{2}\na_a A_b \na^a A^b=-\fr{1}{2}\na_\mu A_\nu \na^\mu
    A^\nu-\fr{1}{2}\na_4 A_\nu \na^4 A^\nu-\fr{1}{2}\na_\mu A_4 \na^\mu
    A^4-\fr{1}{2}\na_4 A_4 \na^4 A^4.
 \en
For 4D observers living in a hypersurface where the fifth
component of the vector field is normal to it, this extra
dimensional field will manifest separately, like an effective 4D
vector field $A^\nu$ and a 4D scalar field $A^4$. In this sense we
can identify kinetic terms for both, scalar and vector fields, and
the derivatives with respect to the extra dimension may be
interpreted as potential (or dynamical sources) terms joined with
massive terms for each of them.\\

 The stress tensor in this gauge is
 \bega
    T_{ab}=&&-\na_aA_e\na_bA^e-\na_e A_a \na^e A_b- 2
    g_{c(b}A_{a)}\Ga^c_{ef}\na^eA^f +\fr{1}{2}g_{ab}\na_e A_f \na^e
    A^f-\\
    \nonumber&&2\fr{g_{,f}}{g}\left[\na_{(a} A_{b)} A^f+\na^f A_{(b} A_{a)}- A_{(a}\na_{b)}
    A^f\right]-
    \left[\na_{(a} A_{b)} A^f+\na^f A_{(b} A_{a)}- A_{(a}\na_{b)} A^f\right]_{,f}.
 \ena

\section{Special case: 5D generalization of a de Sitter spacetime}

Because we are interested to study a cosmological scenario of
inflation from the context of the theory of Space-Time-Matter, we
shall consider the 5D Riemann-flat metric\cite{LB}
 \beg
    dS^2=\psi^2dN^2-\psi^2
    e^{2N}dr^2-d\psi^2, \label{met1}
 \en
where $N$ is a time-like dimension related to the number of
e-folds, $dr^2=dx^i\de_{ij}dx^j$ is the Euclidean line element in
cartesian coordinates and $\psi$ is the space-like extra
dimension. This metric satisfies the vacuum condition $G^{ab}=0$.

For this 5D metric the field equations, after taking Lorentz
gauge: $\na_a A^a=\pa_N A^0+3A^0+\pa_\psi A^4+4\psi^{-1}A^4+\pa_i
A^i=0$, are

 \begin{eqnarray}
   && \left\{\fr{\pa^2}{\pa N^2}+5\fr{\pa}{\pa N}-e^{-2N}\pa^2_r-
    \psi^2\left[\fr{\pa^2}{\pa\psi^2}+\fr{6}{\psi}\fr{\pa}{\pa \psi}\right]\right\}{A^0}
    +\left[\frac{2}{\psi}\fr{\pa}{\pa N}+2 \frac{\pa}{\pa \psi} + \frac{8}{\psi} \right]\,{A^4}=0,\label{b1}\\
   && \left\{\fr{\pa^2}{\pa N^2}+5\fr{\pa}{\pa N}-e^{-2N}\pa^2_r-
    \psi^2\left[\fr{\pa^2}{\pa\psi^2}+\fr{6}{\psi}\fr{\pa}{\pa \psi}
    \right]\right\}{A^j}-2\pa^j\left(A^0+\fr{A^4}{\psi}\right)=0,\label{b2}\\
   && \left\{\fr{\pa^2}{\pa N^2}+3\fr{\pa}{\pa N}-e^{-2N}\pa^2_r -
    \psi^2\left[\fr{\pa^2}{\pa\psi^2}+\fr{6}{\psi}\fr{\pa}{\pa \psi}+
    \fr{12}{\psi^2}\right]\right\} {A^4}=0.\label{b3}
 \end{eqnarray}
Notice that the (\ref{b3}) is decoupled after applying the Lorentz
gauge. However we see that it is not sufficient to decouple all
the field equations. This is because the non zero connections of
the metric (\ref{met1}) act in a non trivial manner in the vector
fields derivatives. There are 14 non zero Christoffel symbols
 \beg
    \Ga^\mu_{\mu4}=\psi^{-1},\ \ \Ga^i_{i0}=1,\ \
    \Ga^0_{ii}=e^{2N},\ \ \Ga^4_{00}=\psi,\ \
    \Ga^4_{ii}=-\psi e^{2N}.
 \en
Therefore, in this Riemann-flat spacetime we obtain the D
'Alambertian of the $A^b$ field
 \beg
    \na_f\na^f A^b=0,
 \en
but, expressed in terms of the ordinary derivatives and the
Christoffel symbols we notice the coupling terms
 \beg
    g^{fh}\left\{\pa_f\pa_h A^b+2\Ga^b_{ef}\pa_h
    A^e+\Ga^b_{he,\,f}A^e-\Ga^e_{fh}\pa_e
    A^b-\Ga^e_{fh}\Ga^b_{ed}A^d+\Ga^b_{ef}\Ga^e_{hd}A^d\right\}=0.
 \en
In the Minkowskian limit ($\psi_0\rightarrow\infty$) all of the
connections vanish and so the field equations remain decoupled
after the gauge choice.

\subsection{Dynamics of the 3D spatially isotropic background fields}

We shall combine the field equations of motion for the classical
homogeneous fields with the Einstein Equations, the first ones
reduce to
 \begin{eqnarray}
    \left\{\fr{\pa^2}{\pa N^2}+5\fr{\pa}{\pa N}-
    \psi^2\left[\fr{\pa^2}{\pa\psi^2}+\fr{6}{\psi}\fr{\pa}{\pa \psi}
    \right]\right\}&{\bar{A}^0}&+\left[\frac{2}{\psi}\fr{\pa}{\pa N}
    +2 \frac{\pa}{\pa \psi} + \frac{8}{\psi} \right]\,{\bar{A}^4}=0,\label{c1} \\
    \left\{\fr{\pa^2}{\pa N^2}+5\fr{\pa}{\pa N}-
    \psi^2\left[\fr{\pa^2}{\pa\psi^2}+\fr{6}{\psi}\fr{\pa}{\pa \psi}
    \right]\right\}&{\bar{A}^j}&=0, \label{c2}\\
    \left\{\fr{\pa^2}{\pa N^2}+3\fr{\pa}{\pa N}-
    \psi^2\left[\fr{\pa^2}{\pa\psi^2}+\fr{6}{\psi}\fr{\pa}{\pa \psi}+
    \fr{12}{\psi^2}\right]\right\}&{\bar{A}^4}&=0. \label{c3}
 \end{eqnarray}
Notice that the equation for $\bar A^0$ is the unique coupled.
Furthermore, once obtained ${\bar{A}^4}$, we can describe the
dynamics of ${\bar{A}^0}$ in (\ref{c1}), where ${\bar{A}^4}$
appears as a source.

\section{Effective 4D dynamics of the fields}

The remarkable property of the 5D metric (\ref{met1}) is that it
is a generator of 4D de Sitter spacetimes. This may be done when
we foliate the space (\ref{met1}) in a particular hypersurface
$\psi=\psi_0$. It is said that for an observer moving with the
penta velocity $U_\psi=0$, the spacetime describes a de Sitter
expansion. Then the effective 4D hypersurface it has a scalar
curvature $^{(4)} R=12/\psi^2_0=12\,H^2_0$, such that the Hubble
parameter is defined by the foliation $H_0=\psi_0^{-1}$. If we
consider the coordinate transformations
 \beg\label{trans}
    t=\psi_0N,\ \ \ R=\psi_0 r,\ \ \ \psi=\psi,
 \en
we then arrive to the Ponce Leon metric\cite{PdL}:
$dS^2=\left(\fr{\psi}{\psi_0}\right)^2\left[dt^2-e^{2t/\psi_0}dR^2\right]-d\psi^2$.
If we foliate $\psi=\psi_0$, we get the effective 4D metric
 \beg
    dS^2\rightarrow ds^2=dt^2-e^{2H_0t}d\vec{R}^2,\label{24}
 \en
which describes a 3D spatially flat, isotropic and homogeneous de
Sitter expanding universe with a constant Hubble parameter $H_0$.

The dynamics of the fields being given by the equations
(\ref{b1}), (\ref{b2}) and (\ref{b3}), evaluated on the foliation
$\psi=\psi_0=1/H_0$, with the transformations (\ref{trans}). In
the following subsections we shall study separately the dynamics
of the classical 3D spatially isotropic fields:
$\bar{A}^{\mu}(t,\psi_0)$ and $\bar{A}^{4}(t,\psi_0)$, and the
fluctuations of these fields: $\delta A^{\mu}(t,\vec{R},\psi_0)$
and $\delta A^{4}(t,\vec{R},\psi_0)$. Notice that now
$\vec{R}\equiv \vec{R}(X^i)$. To describe the dynamics of the
fields we shall impose the effective 4D Lorentz gauge:
$^{(4)}\nabla_{\mu} A^{\mu}=0$. It implies that the 5D Lorentz
gauge with the transformations (\ref{trans}) and evaluated on the
foliation must now be
\begin{equation}\label{ga}
\left.\na_a A^a\right|_{\psi_0} =^{(4)} \nabla_{\mu}
A^{\mu}(t,\vec{R},\psi_0) +\left.\left(\pa_\psi
A^4+4\psi^{-1}A^4\right)\right|_{\psi_0}=0,
\end{equation}
where $^{(4)} \nabla_{\mu}\,A^{\mu}$ denotes the covariant
derivative on the effective 4D metric (\ref{24}). Hence, in order
to the effective 4D Lorentz gauge to be fulfilled, we shall
require
\begin{equation}\label{gau}
\left.\left(\pa_\psi A^4+4\psi^{-1}A^4\right)\right|_{\psi_0}=0.
\end{equation}

\subsection{4D classical field dynamics}\label{cfd}

In order to solve the equations (\ref{c1}), (\ref{c2}) and
(\ref{c3}) on an effective 4D de Sitter metric, we must evaluate
these equations on the particular foliation
$\psi=\psi_0=H^{-1}_0$, $r=R\,\psi_0$ and $N=H_0\,t$. We shall
identify the effective scalar $A^4$ with the inflaton field:
$A^4(t,\vec R,\psi_0)\equiv
  \phi(t,\vec R,\psi_0)$ and we shall denote $\bar\phi(t,\psi_0)
\sim \left.\phi_1(N)\,\phi_2(\psi)\right|_{N=H_0
t,\psi=\psi_0=H^{-1}_0}$, as the 3D spatially isotropic and
homogeneous background field. In the same way we state for the
homogeneous component of the vector field the separation $\bar
A^j(t,\psi_0)\sim \left.S^j_{1}(N)S^j_{2}(\psi)\right|_{N=H_0
t,\psi=\psi_0=H^{-1}_0}$, in the next we shall drop the index $j$
to label the functions $S_1(t)$ and $S_2(\psi_0)$. Hence, we
obtain
 \beg
    \bar\phi(t,\psi_0)\sim \phi_{m}(t,y_0)=e^{-\fr{3}{2}H_0 t}\left(a_1 \, e^{\al H_0 t}+a_2 \, e^{-\al
    H_0 t}\right),\ \ \ \ \al=\fr{3}{2}\sqrt{1-\fr{4m^2}{9}},
 \en
 where we have considered the condition (\ref{gau}), such that
 \beg
\left. -\psi^2 \left[ \frac{\pa^2}{\pa\psi^2} + \frac{3}{\psi}
 \frac{\pa}{\pa\psi} \right] \bar{A}^4\right|_{\psi_0} = m^2
 \,\bar{\phi}(t,\psi_0).
 \en
 Furthermore, the general solution of eq. (\ref{c2}) on the effective 4D metric
 (\ref{24}), is
  \beg
    \bar
A^j(t,\psi_0)\sim S_{\mu}(t)= e^{-\fr{5}{2}H_0 t}\left(c_1 \,
e^{\ga H_0 t}+c_2 \, e^{-\ga
    H_0 t}\right),\ \ \ \ \ga=\fr{5}{2}\sqrt{1-\fr{4\nu^2}{25}}
    \en
where \beg \left. -\psi^2 \left[ \frac{\pa^2}{\pa\psi^2} +
\frac{6}{\psi}
 \frac{\pa}{\pa\psi} \right] \bar{A}^j\right|_{\psi_0} = \nu^2
 \,\bar{A}^j(t,\psi_0).
\en A similar treatment can be done for $\bar{A}^0$, after making
use of the condition (\ref{gau}), the transformations
(\ref{trans}) and the foliation $\psi=\psi_0=1/H_0$. However, the
difference with the other background components of the field
observed in eq. (\ref{c1}) is that $\bar{A}^4\equiv
\bar\phi(t,\psi_0)$ acts as a source of $\bar{A}^0(t,\psi_0)$.

As a particular choice we shall consider a 4D inflationary
universe, where the background fields are $\bar
A^b=\left(0,0,0,0,\bar\phi\right)$, in agreement with a global (de
Sitter) accelerated expansion which is 3D spatially isotropic,
flat and homogeneous\footnote{One could consider, for instance,
the case when the background field is $\bar A^b=\left(\Phi,\bar
A^1,0,0,0\right)$, that defines an effective homogeneous component
of the electric field. However, we would obtain an anisotropic
component of the stress tensor $T_{10}$, which is not compatible
with our background, spatially flat, homogeneous and {\em
isotropic} (de Sitter) metric. In general this implies that for
the background fields to satisfy Einstein equations, the
components $\bar A_0;\bar A_1;\bar A_2;\bar A_3$ are highly
restricted. In particular we have the following cases to
choose:(i)$\bar A^i=0$, $\bar A^0=\bar\Phi(t,\psi_0)$ and $\bar
A^4=\bar\phi(t,\psi_0)$, (ii)$\bar A^0=\bar A^4=0$ and $\bar
A^i=\bar A_0^i$ constants. In what follows we shall analyze a
particular choice of the first case (with $\bar A^0=0$), because
the other isn't very interesting in the physical sense.}.
 In this case, the relevant
components of the classical Energy momentum tensor, are
 \bega
    \rho\equiv\langle T^0_0\rangle&=&\fr{1}{2}\dot{\bar\phi}^2+
    \left[\fr{5}{\psi^2} \bar\phi^2+ \fr{1}{2} \bar{\phi}'^2+\fr{2}{\psi}
    \bar\phi \bar{\phi}'\right]_{\psi=\psi_0}, \label{t1}\\
    p\equiv\langle
    -T^i_j\rangle|_{i=j}&=&\fr{1}{2}\dot{\bar\phi}^2-
    \left[\fr{5}{\psi^2} \bar\phi^2+ \fr{1}{2} \bar{\phi}'^2+\fr{2}{\psi}
    \bar\phi \bar{\phi}'\right]_{\psi=\psi_0}, \label{t2}\\
    \langle T^\al_\be\rangle|_{\al\neq\be}&=&0, \label{t3}
 \ena
where dots denote derivative with respect to the time which in our
case are zero: $\left.\dot{\bar\phi}\right|_{\psi_0}=0$.
Furthermore, from eq. (\ref{t1}) we can make the following
identification for the background scalar potential:
\begin{equation}
V[\bar\phi]=\left[\fr{5}{\psi^2} \bar\phi^2+ \fr{1}{2}
\bar{\phi}'^2+\fr{2}{\psi} \bar\phi
\bar{\phi}'\right]_{\psi=\psi_0}.
\end{equation}
In our model, the hypersurface $\psi=\psi_0$ defines a de Sitter
expansion of the universe with a Hubble parameter
$H_0=\psi_0^{-1}$. The equation of state for this case is
$p=-\rho=-3/\left(8 \pi G\psi^2_0\right)$. Then, it is easy to see
that the only compatible background solution for the field
evaluated on the hypersurface is the typical de Sitter solution
for a background scalar field: $\bar\phi(t,\psi_0)=\bar\phi_0$.
This means that
\begin{equation}\label{pot}
V[\bar\phi_0]=\left[\fr{5}{\psi^2} \bar\phi^2_0+ \fr{1}{2}
\bar{\phi}'^2+\fr{2}{\psi} \bar\phi_0
\bar{\phi}'\right]_{\psi=\psi_0}=\frac{3 H^2}{8\pi G}.
\end{equation}
A particular solution of (\ref{pot}) is
\begin{eqnarray}
&& \left.\left(\bar\phi'\right)^2\right|_{\psi=\psi_0=1/H} = \frac{5 H^2}{4\pi G}, \label{e1}\\
&& \bar\phi' = - 5 H \bar\phi. \label{e2}
\end{eqnarray}
From eqs. (\ref{e1}) and (\ref{e2}) we obtain
\begin{equation}
\left.\left(\bar\phi\right)^2\right|_{\psi=\psi_0=1/H} =
\bar\phi^2_0=\frac{1}{20\,\pi\,G}.
\end{equation}

\subsection{4D Field fluctuations}

Here we consider equations (\ref{b1}), (\ref{b2}) and (\ref{b3})
to search for possible electromagnetic fields generated through
this model. In Sect. (\ref{cfd}) we've seen that the Einstein
equations for the background fields exclude any possibility of
homogeneous electromagnetic fields.

The equation for the effective scalar $\delta
A^4(t,\vec{R},\psi_0)$ on the effective hypersurface (\ref{24}) is
decoupled from the dynamics of the 4-vector. In contrast, the
equations for $\de A^0(t,\vec{R},\psi_0)$ and $\de
A^i(t,\vec{R},\psi_0)$ remain coupled. By the use of our 5D
Lorentz gauge evaluated on the foliation $\psi=\psi_0=H^{-1}_0$:
$\left. \nabla_{a} \,A^a\right|_{\psi_0=H^{-1}_0}=0$, we can
express the inhomogeneous term for $\de A^0$ as only a function of
$\de A^4$. The solution will involve both, homogeneous and
inhomogeneous parts. Once obtained $\delta A^0$ and $\delta A^4$,
we can finally search solutions for the components $\de A^j$.
These total solutions are necessary to deduce the effective
electric fields. In contrast, as we previously said, the equation
of motion for pure magnetic fields may be obtained by just
applying the curl in the 3-space to equation (\ref{b2}). The last
term in (\ref{b2}) vanishes because is a 3-gradient, and so
magnetic fields equations are decoupled. To quantize the field
fluctuations on the effective 4D de Sitter spacetime (\ref{24}),
we shall consider the equations (\ref{b1}), (\ref{b2}) and
(\ref{b3}), with condition (\ref{gau}), the transformations
(\ref{trans}) and the foliation $\psi=\psi_0=1/H_0$. The equal
time canonical relations are

 \beg\label{com1}
  \left.\left[ \de A_i(t,\vec{R},\psi_0),
  \Pi^j(t,\vec{R'},\psi_0)\right]\right|_{\psi_0=1/H_0} = -i\,
  g^j_i e^{-3H_0t}\,\delta^{(3)}(\vec{R} - \vec{R'}),
 \en
where
$g^{ij}$ are the space-like components of the tensor metric in
(\ref{24}) and $\delta^{(3)}(\vec{R} - \vec{R'})$ is the 3D
Dirac's function. Furthermore, the canonical momentum is given by
the electric field $\Pi^j \equiv E^j= \nabla^j A^0 - \nabla^0
A^j$. The equations (\ref{b1}), (\ref{b2}) and (\ref{b3}) with the
transformations (\ref{trans}) can be evaluated on the foliation
$\psi=\psi_0=1/H_0$ to give the dynamics on the effective 4D
spacetime (\ref{24}). If we take into account the conditions
(\ref{gau}), the effective 4D dynamics of the fluctuations
describe an effective 4D Lorentz gauge, so that
\begin{eqnarray}
&&\frac{\partial^2\delta A^0}{\partial t^2} + 5 H_0
\frac{\partial\delta A^0}{\partial t} - H^2_0 e^{-2H_0 t}
\partial^2_R \delta A^0 + \nu^2H^2_0 \delta A^0= -2 H^2_0
\frac{\partial\delta
\phi}{\partial t}, \label{f1} \\
&&\frac{\partial^2\delta A^j}{\partial t^2} + 5 H_0
\frac{\partial\delta A^j}{\partial t} - H^2_0 e^{-2H_0 t}
\partial^2_R \delta A^j + \nu^2H^2_0 \delta A^j= 2 H^2_0
\pa^j\left(\delta A^0+ H_0 \delta\phi\right),
\label{f2} \\
&&\frac{\partial^2\delta\phi}{\partial t^2} + 3 H_0
\frac{\partial\delta\phi}{\partial t} - H^2_0 e^{-2H_0 t}
\partial^2_R \delta\phi + m^2 H^2_0 \delta\phi= 0. \label{f3}
\end{eqnarray}
describe the 4D dynamics of the fluctuations. A very important
fact is that the electromagnetic field fluctuations $\delta
A^{\mu}$ obey a Proca equation with sources. The expansion of the
field in temporal modes is
 \beg
    \de A^\mu(t,\vec{R},\psi_0)=\int \fr{d^3 K}{(2\pi)^3}\sum_{\la=1}^3
    \varepsilon^\mu(\vec{K},\la)\left(a_{(\vec{K},\la)}e^{-i\vec{K}\cdot\vec{R}}S(K,t,\psi_0)+
    a^\dagger_{(\vec{K},\la)}e^{i\vec{K}\cdot\vec{R}}S^\star(K,t,\psi_0)\right),
    \en
where $\vec K= H_0 \, \vec k$ ($k$ is a dimensionless wavenumber).
Furthermore, $\varepsilon^{\mu}(\vec{k},\lambda)$ are the
polarizations \footnote{parenthesis denotes that sum do no run
over these indices.}, such that in the Lorentz gauge the following
expression holds:
\begin{equation}
\sum_{\lambda=1}^{3}
\varepsilon_{\alpha}(\vec{k},\lambda)\,\varepsilon_{\beta}(\vec{k},\lambda)
= -\left(g_{\al\be}-\fr{H_0^2}{m^2_{eff}}k_\al k_\be\right),
\end{equation}
where we have introduced the effective mass
$m^2_{eff}=H_0^2(\nu^2-\fr{25}{4})$ of the redefined temporal
modes ${\cu}_K(t)=e^{5H_0t/2}S(K,t,\psi_0)$, that obey the
harmonic equation $\ddot\cu_K+\omega^2_K(t)\cu_K=0$. The time
dependent frequency is defined by the relation $K_\mu
K^\mu=m^2_{eff}$.
 \beg
    \omega^2_K(t)=\left[m^2_{eff}+(e^{-H_0t} K)^2\right].
 \en
 Modes with $\omega^2_K >0$ are stable, but those with $\omega^2_K
 <0$ [i.e., with $k < \left(25/4 - \nu^2 \right)^{1/2} e^{H_0 t}$], are unstable.
In the small wavelength limit these behave like plane waves in
Minkowski space. Furthermore, the annihilation and creation
operators $a_{(K,\lambda)}$ and $a^{\dagger}_{(K,\lambda)}$,
comply with the commutation relations
\begin{equation}
\left[a_{(\vec{K},\lambda)},
a^{\dagger}_{(\vec{K}',\lambda')}\right] = \left(2\pi\right)^3
g_{\lambda\lambda'}\,\delta^{(3)}(\vec{K} - \vec{K}').
\end{equation}

The time dependent modes for the contravariant vector $\de A^\mu$
are
 \begin{equation}
S(K,t,\psi_0)
=e^{-5H_0t/2}\left\{c_1\ch^{(1)}_\sigma\left[x(t)\right]+c_2\ch^{(2)}_\sigma\left[x(t)\right]\right\},\
\ \ \ \
                               \sigma=\sqrt{\fr{25}{4}-\nu^2},\ \ \ \ \ x(t)={K\over H_0}\, e^{-H_0t}.\\
 \end{equation}
We can also obtain the temporal modes for the covariant $\de
A_\mu$ which are related to the contravariant ones:

$\ct_K(t)=e^{2H_0t}\,S(K,t,\psi_0)$. The commutation relations
(\ref{com1}) yield the following conditions over these modes
 \bega
    \ct_K\dot\ct_K^\star-\ct_K^\star\dot\ct_K&=&-i e^{-H_0t}, \label{e47}\\
    \ct_K\ct_K^\star&=&\fr{e^{-H_0t}}{2w_K(t)}, \label{e48}
 \ena
which are only valid on short wavelength modes for which
$\omega^2_K >0$. Equations (\ref{e47}) and (\ref{e48}) give us the
normalization conditions for the modes of $\delta A_{\mu}$. On the
other hand, these modes are unstable on cosmological scales:
$\omega^2_K <0$, and the expression (\ref{e47}) tends to zero. To
apply these conditions we take the small wavelength limit for the
Hankel Functions $x(t)\gg|\sigma^2-\fr{1}{4}|$. These means that
$K/H_0\,e^{^-H_0t}\gg m_{eff}^2$, so that $w_K(t)\simeq K
e^{-H_0t}$. In this limit the conditions (\ref{e47}) and
(\ref{e48}) become dependent one of the another. If we choose
$c_1=0$, the solution for the modes is
 \beg
    \ct_K(t)=e^{-\fr{1}{2}H_0t}\sqrt{\fr{\pi}{4H_0}}\,\ch^{(2)}_\sigma[x(t)],
 \en
 where $\ch^{(2)}_\sigma[x(t)]$ is the second kind Hankel
 function.

\subsubsection{4D electromagnetic fluctuations}

The electric field for a observer in 4D is defined by its
4-velocity $E_\nu=F_{\nu\la}u^\la$. If we choose the particular
co-moving frame $u^\nu=\left[(H_0\psi_0)^{-1},\vec{0}\right]$, we
obtain
 \bega
    \nonumber E_0&=&0,\\
              E_i&=& \fr{\pa}{\pa
              X^i} \de A^0- e^{2H_0 t} \fr{\pa }{\pa
              t} \de A^i-2 H_0 \,e^{2H_0 t} \de A^i.
 \ena
The magnetic fields are defined by
$B_\nu=\fr{1}{2}\eps_{\nu\la\al\be}u^\la F^{\al\be}$, where
$\eps_{\nu\la\al\be}=\sqrt{\left|{}^{(4)}g\right|}{\cal
A}_{\nu\la\al\be}$ is the totally antisymmetric Levi-Civita tensor
and ${\cal A}_{\nu\la\al\be}$ is a totally antisymmetric symbol
with ${\cal A}_{0123}=-1$. Then for a co-moving observer we will
have a magnetic field,
 \bega
\nonumber    B_0&=&0,\\
\nonumber    B_j&=&\fr{\sqrt{\left|{}^{(4)}g\right|}}{2}\,{\cal
                    A}_{j0kl}\,u^0\,F^{kl}.
 \ena
From the last expression we can arrive to another that will be
useful to obtain an equation of motion for the magnetic fields, we
first define the Levi-Civita symbol in the 3-flat space using the
co-moving frame: $\eps_{jkl}={\cal A}_{j0kl}$ (we note that
$\eps_{123}=1$). Hence
 \beg
    B_j=\sqrt{\left|{}^{(4)}g\right|}
    g^{kk'}u^0\eps_{jkl}\pa_{k'}A^l.
 \en
For our particular case we obtain
 \beg
  e^{-H_0 t}\, B_j=\left[\delta^{k k'} \eps_{jkl}\, \pa_{k'}\right]
   A^l.
 \en
The differential operator between square brackets commutes with
the one applied to $A^j$ in the equation (\ref{b2}), so that in
the equation of motion for ${\cal B}_j=e^{-H_0t}B_j$ there are no
sources. We can express the field in Fourier components of the
$\de A^j$ field
 \bega
    \cB^j\left(t,\vec{R},\psi_0\right)=\int
    \fr{d^3K}{(2\pi)^3}\sum_{\la=1}^{3}\varepsilon_{l}(\vec{K},\la)\eps^{jnl}\left[ a_{(\vec{K},\la)}
    {\cv}_{n}(K,t,\psi_0)\,
    e^{i\vec{K}\cdot\vec{R}} + a^{\dagger}_{(\vec{K},\la)} \cv^\star_{n}(K,t,\psi_0)\,e^{-i\vec{K}\cdot\vec{R}}\right].
 \ena
Here $\cv_{j}(K,t,\psi_0)=-iK_j\, S_{1}(K,t,\psi_0)$ are the
temporal modes with their complex conjugate
$\cv^\star_{j}(K,t,\psi_0)=iK_j\,S^\star(K,t,\psi_0)$. We perform
the vacuum expectation value of the B-fields quadratic amplitude,
defined by the invariant product $\langle B^2\rangle\equiv
\langle0|B^\al B_\al|0\rangle$. For comoving observers $B^0=0$ and
so we have $B^2=B^j\,B_j=e^{-2H_0t}\sum_j{B_j}^2=\sum_j{\cB_j}^2$.
Then
 \beg
    \langle B^2\rangle=\int \fr{d^3K}{(2\pi)^3}(2e^{2H_0t}K^2)S(K,t,\psi_0)
    S^\star(K,t,\psi_0).
 \en
We will cut the above integral up to wavelengths that remain well
outside the horizon wavenumber $k_H=\fr{5}{2}e^{H_0t}$. In this
limit we use the asymptotic limit of the hankel functions for the
long wavelength limit $k\,e^{-H_0t}\ll\sqrt{\sigma+1}$. The power
spectra is then
 \beg
    {\cal
    P}_B(k)=\fr{2^{2\sigma}\Ga^2(\sigma)H_0^4}{4\pi^3}\,e^{(2\sigma-3)H_0t}\,k^{5-2\sigma},
 \en
if we ask for an almost scale invariant spectrum, then
$\sigma=\fr{5}{2}+\eps,\ \ \eps=-\fr{\nu^2}{5}$ and $\nu^2\ll1$.
The quadratic amplitude is then
 \beg
    \langle
    B^2\rangle=\fr{45H_0^4}{4\pi^2\nu^2}e^{2H_0t}\left(\fr{5\theta}{2}\right)^{-2\eps},
 \en
where $\theta \ll 1$ is a control parameter, such that we stay
with super Hubble wavelenghts: $k < \theta\, k_H$.

Using the homogeneous solutions of the equations (\ref{f1}),
(\ref{f2}) and (\ref{f3}) we can deduce their contribution for
electric fields on the infrared (IR) sector, we obtain for
comoving observers $\langle E^2\rangle_{IR}=\langle
E_A^2+E_B^2+E_C^2\rangle_{IR}$, where
 \bega
\langle E_A^2\rangle_{IR} &\simeq&-H_0^5\fr{e^{-4H_0t}}{\left(\nu^2-\fr{25}{4}\right)}\,
\int_0^{\theta k_H}\fr{dk}{2\pi^2}\,k^6|\ct_k|^2,\\
\langle E_B^2\rangle_{IR} &\simeq&-H_0^5\,e^{-2H_0t}\,\int_0^{\theta k_H}\fr{dk}{2\pi^2}\,
\left(3\,e^{2H_0t}+\fr{H_0^2k^2}{m_{eff}^2}\right)|\dot\ct_k|^2,\\
\langle E_C^2\rangle_{IR} &\simeq & H_0^5\,e^{-2H_0t}\,
\int_0^{\theta
k_H}\fr{dk}{2\pi^2}\sum_j\fr{H_0^2k_0k_j}{m_{eff}^2}(-iH_0k_j)\left(\ct_k\dot\ct_k^\star-\ct_k^\star\dot\ct_k\right).
 \ena
If we choose $\sigma=\fr{5}{2}+\eps,\ \ \eps=-\fr{\nu^2}{5}$ and
$\nu^2\ll1$, we get
 \bega
 \langle E_B^2 \rangle_{IR} &\simeq &
 \left(\fr{9}{5\pi}\right)^2\,H_0^4\,e^{2H_0t}\left[3\theta^{-2}+\fr{4}{25}\theta^{-2\eps}\right],
 \\
  \langle E_B^2 \rangle_{IR} & \simeq&3\left(\fr{9}{5\pi}\right)^2\,H_0^4\,e^{2H_0t}\theta^{-2},\\
  \langle E_C^2 \rangle_{IR} &\simeq &0,
 \ena
on cosmological scales. Notice that $\langle E^2 \rangle$ is not
scale invariant. Then we can say that on very large scales the
amplitude of electromagnetic fields are
 \bega
    \left<B^2\right>^{1/2}_{IR}\simeq\fr{3\sqrt5}{2\pi\nu}H_0^2
    e^{H_0t}\left(\fr{5\theta}{2}\right)^{\nu^2/5},\ \ \ \ \
    \left<E^2\right>^{1/2}_{IR}\simeq\fr{3^{5/2}}{5\pi}H_0^2e^{H_0t}\theta^{-1}.
 \ena

\subsubsection{4D inflaton fluctuations}

For the fluctuations of the inflaton field we can make a similar
treatment. The Fourier expansion is
 \beg
    \de \phi\left(t,\vec{R},\psi_0\right)=\int
    \fr{d^3K}{(2\pi)^3}
\left[ \alpha_{(\vec{K})} \phi(K,t,\psi_0)\,
e^{i\vec{K}\cdot\vec{R}} + \alpha^{\dagger}_{(\vec{K})}
\phi^*(K,t,\psi_0)\,e^{-i\vec{K}\cdot\vec{R}}\right],
 \en
such that the annihilation and creation operators
$\alpha_{(K,\lambda)}$ and $\alpha^{\dagger}_{(K,\lambda)}$,
comply with the commutation relations
\begin{equation}
\left[\alpha_{(\vec{K})}, \alpha^{\dagger}_{(\vec{K}')}\right] =
\left(2\pi\right)^3 \,\delta^{(3)}(\vec{K} - \vec{K}').
\end{equation}
The solutions for the modes $\phi(K,t,\psi_0)$, are
 \begin{equation}
    \phi(K,t,\psi_0)  = e^{-3H_0 t/2}\,\left\{c_1\,J_\mu\left[x(t)\right]
    +c_2\,Y_\mu\left[x(t)\right]\right\},\ \ \ \ \ \mu=\sqrt{\fr{9}{4}-m^2}.
 \end{equation}
The nearly invariant spectrum of the scalar perturbations is
obtained for small values of the effective mass: $m\ll 1$. After
normalization of the modes, we obtain the standard result(see, for
instance\cite{bcms}) on cosmological scales
\begin{equation}
\left< \delta\phi^2\right>_{IR} \simeq
\frac{\Gamma^2(\mu)}{\pi^3(3-2\mu)}\left(\frac{2
}{\theta\mu}\right)^{2\mu-3}\,  H^2_0,
\end{equation}
which is divergent for an exactly scale invariant power spectrum
corresponding to a null value of $m$.

\section{Final Comments}

We have shown how primordial electromagnetic fields and inflaton
fluctuations can be generated jointly during inflation using a
semiclassical approach to GEMI. The difference with respect other
previous works is that, in this one, we have used a Lorentz gauge
(rather than a Coulomb gauge). One of the important facts is that
our formalism is naturally not conformal invariant on the
effective 4D metric (\ref{24}), which make possible the super
adiabatic amplification of the modes of the electromagnetic fields
during inflation in a comoving frame on cosmological (super
Hubble) scales.

In this letter we have analyzed the simplest nontrivial
configuration field:
$\bar{A}^b=\left[0,0,0,0,\bar\phi(t,\psi_0)\right]$. For these
configuration background fields to satisfy the Einstein equations
in a de Sitter expansion, the background inflaton field must be a
constant on the metric (\ref{24}):
$\bar\phi(t,\psi_0)=\bar\phi_0$. Then, in the model here
developed, the expansion of the universe is driven by the
background inflaton field $\bar\phi_0$ and background
electromagnetic fields are excluded to preserve global isotropy.
Notice that back reaction effects are not included in the
semiclassical approach here used for the treatment of the Einstein
equations. These effects should be included jointly with vectorial
metric fluctuations and are the subject of a future work.

To describe the dynamics of the fields, we impose the effective 4D
Lorentz gauge $^{(4)}\nabla_{\mu} A^{\mu}=0$, given simultaneously
by conditions (\ref{ga}) and (\ref{gau}). Therefore, the origin of
the generation of the seed of electromagnetic fields and the
inflaton field fluctuations during inflation can be jointly
studied. The dynamics of $\delta A^{\mu}$ on the effective 4D
metric (\ref{24}) obey a Proca equation with sources where the
effective mass of the electromagnetic field fluctuations is
induced by the foliation $\psi=\psi_0=1/H_0$. From the point of
view of a relativistic observer this foliation imply that the
component of the penta-velocity $U^{\psi} ={d \psi\over d S}=0$.

Finally, we obtain for small values of the mass $\nu$ a nearly
scale-invariant long wavelengths power spectrum for $\left<
B^2\right>$, which grows as $a^2$ during inflation. After
inflation these fields decreases as $a^{-2}$ to take the present
day values on cosmological scales: $\left.\langle B^2\rangle
^{1/2}\right|_{{\rm Now}} < 10^{-9}$ Gauss\cite{ul}. On the other
hand, using the homogeneous solutions for $\ct_k(t)$, we obtained
that $\left< E^2\right>_{IR}$ also grows as $a^2$ but has a scale
dependent power spectrum that goes as ${\cal A}_1(t)\,k^2+{\cal
A}_2(t)\,k^{-2}$; the first term means that (for a given time),
electric fields become more important on shorten scales and the
second one become more important on very large scales. In what
respect to the inflaton field fluctuations $\left<
\delta\phi^2\right>$, they are scale invariant on cosmological
scales, but the amplitude is freezed in agreement with the
predictions of standard 4D inflation.

\begin{acknowledgments}
The authors acknowledge CONICET and UNMdP (Argentina) for
financial support.
\end{acknowledgments}

\end{document}